\documentclass[aps,prb,twocolumn]{revtex4}

\usepackage{amssymb}
\usepackage{amsmath}
\usepackage{graphicx}

\begin{document}
\title{Observation of large $h/2e$ and $h/4e$ oscillations \\in a proximity dc superconducting quantum interference device}

\author{J. Wei, P. Cadden-Zimansky and V. Chandrasekhar}
\affiliation{Department of Physics and Astronomy, Northwestern University, Evanston, IL 60208, USA}

\begin{abstract}
We have measured the magnetoresistance of a dc superconducting quantum interference device in the form of an interrupted
mesoscopic normal-metal loop in contact with two superconducting
electrodes. Below the transition temperature of the
superconducting electrodes, large $h/2e$ periodic
magnetoresistance oscillations are observed. By adding a small dc
bias to the ac measurement current, $h/4e$ oscillations can be
produced. Lowering the temperature further leads to even larger
oscillations, and eventually to sharp switching from the
superconducting state to the normal state. This flux-dependent resistance could be utilized to make highly sensitive flux detector.
\end{abstract}
\maketitle

Superconducting Quantum Interference Devices (SQUIDs) are ideal
for detecting extremely small changes in magnetic fields and have
undergone extensive development and evolution for over 40
years.~\cite{Barone1982,Clarke2004} The dc SQUID consists of two
Josephson junctions connected in parallel. When biased above its
critical current, the dc voltage across the SQUID is modulated by
a magnetic flux. For proper operation, additional resistive shunts
across the junctions are usually required to remove the hysteretic
behavior of the device. The Nyquist noise of these resistive
shunts limits the sensitivity of the SQUID. Recently, with
advances in nanofabrication techniques, it has become possible to
fabricate a mesoscopic SQUID with the Josephson junctions composed
of normal metal sections with lengths shorter than the normal
metal coherence length $\xi_{\text{N}}$.  The use of normal metal
junctions enables the SQUID to be intrinsically shunted, allowing
for the possibility of increased sensitivity. Additionally, it may
eliminate the low frequency noise due to weakly trapped charges in the
Josephson barrier, which is a major obstacle for realizing
long-lived quantum states in SQUIDs.~\cite{Muck2005} Since the
resistance of the superconducting-normal-superconducting (SNS) junction changes with temperature due to the
superconducting proximity effect, the properties of this proximity
dc SQUID are quite different from traditional SQUIDs which use
junctions made of thin insulating barriers. In this work, we
investigate the behavior of such a proximity dc SQUID and find
unusually large magnetoresistance oscillations with both $h/2e$
and $h/4e$ periods. In addition, at low temperatures the SQUID
undergoes flux-dependent switching from the superconducting to the
normal state, which may be useful for applications.

The inset of Fig.~1 shows SEM images of two different devices. The
geometry is that of an interrupted normal-metal Au loop in contact with two
superconducting Al electrodes. The loop is broken into two arms,
which each serve as a Josephson junction between the
superconducting electrodes. The separation between the two arms is
100 nm for the upper device in the inset, and 40 nm for the
lower one (the separation is not clear in this SEM image due to the
overlap of the Al and Au layers). This separation is smaller than
the typical superconducting coherence length $\xi_{\text{S}}$ for
Al.~\cite{Cadden-zimansky2006}  The width of the normal metal arms
of the loop is 90 nm and the length of each arm is approximately 1
$\mu$m. As shown by the schematic in Fig.~1, the superconducting
Al electrodes extend for several micrometers on each side before
overlapping with Au leads used to make four-probe measurements.
Conventional bilayer (PMMA/MMA) $e$-beam lithography was used to
pattern the devices on oxidized Si substrates. The 50 nm thick Au
was deposited first in a thermal evaporator with base pressure
$3\times 10^{-7}$ Torr at a rate of 0.6 nm/sec. After further
patterning, an \textit{in situ} Ar$^+$ plasma etch was performed
just prior to the deposition of the 80 nm thick Al wires to ensure
transparent interfaces. After the fabrication of the Al wires the
devices were immediately loaded into a dilution refrigerator with
base temperature lower than 20 mK. Differential resistance
measurements were performed with conventional ac bridge and
lock-in techniques using measurement currents of 20 -100 nA.

\begin{figure}
\includegraphics[width=3.1in]{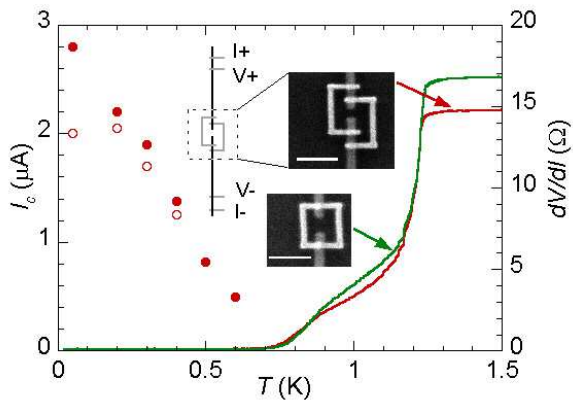}
\caption{ (Color online) Right scale: Resistance as a function of temperature for
the two proximity SQUIDs shown in the inset; the data are taken
using an ac measurement current of \textit{I}$_{\text{meas}}$ = 100 nA.
Left scale: Temperature dependence of the critical current derived
from differential resistance vs. bias current measurements for the
upper one of the two devices. The ac measurement current is 100 nA
above 450 mK and 20 nA below 450 mK. Below 450 mK, the retrapping
currents (open circles) were clearly different from the switching
currents (solid circles), indicating hysteresis. Inset: Schematic
of the four-probe measurement configuration and SEM images of the
two devices. The arms of the loops (brighter wires) are made of Au
and the electrodes extending out of the top and bottom (darker
wires) are made of Al. The size bar in both images is 500 nm.}
\end{figure}

The resistance for each of the two devices as a function of
temperature is shown in Fig.~1. Since both devices yielded similar
results, in what follows we will focus our discussion on the upper
one. As the temperature is decreased there is a sharp drop in the
resistance at 1.2 K, corresponding to the transition temperature
of the Al electrodes between the voltages probes. The remaining
resistance of about 5 $\Omega$  just below this transition
temperature is the resistance of the Au loop. Due to the proximity
effect caused by the superconducting Al electrodes, this
resistance drops as the temperature is further decreased,
eventually reaching zero at 0.7 K. The pronounced nature of the
proximity effect just below 1.2 K is an indication of highly
transparent N-S interfaces.~\cite{Courtois1996} Below 0.7 K, where
the resistance is zero, the critical current of the loop is
measured by applying a dc bias current in addition to the ac
measurement current. As shown in Fig.~1, the critical current
continues to increase with decreasing temperature and does not
saturate. However, below 0.45 K the critical current displays
hysteresis depending on the direction of the current sweep, as has
been reported in other experiments on similar
structures.~\cite{Hoffmann2004, Angers2007} Whether the hysteresis
is due to thermal effects or intrinsic damping caused by the
normal metal has not been determined.~\cite{Angers2007}

From the normal resistance of the loop, and the resistance of a
simultaneously fabricated long normal wire sample on the same
chip, the resistance per square $R_{\square}$ of the Au film is found to be
 0.8 $\Omega$. From this we calculate an elastic mean
free path $l=21$ nm and a diffusion constant \textit{D} = 97
cm$^{2}$/sec. Since the length of each normal metal arm $L$ is
approximately 1 $\mu$m, the corresponding Thouless energy,
$\epsilon_{\text{c}}=\hbar D/L^{2}$, is 6.6 $\mu$eV and Thouless
temperature, $T_{\text{Th}}=\epsilon_{\text{c}}/k_{B}$, is 77 mK.
For SNS proximity junctions, theory predicts that in the long
junction limit, i.e., $L\gg \sqrt{\hbar D/\Delta}$ or
$\Delta/\epsilon_{\text{c}} \gg 1$, the critical current increases
exponentially with decreasing $T$, and as $T$ approaches
$T_{\text{Th}}$ the ratio $eR_{\text{N}}I_{\text{c}}/\epsilon_{c}$
should saturate at 10.82.~\cite{Dubos2001a} However, for our dc
SQUID with two junctions in parallel, $\Delta/\epsilon_{\text{c}}
\sim 26$, but the ratio
$eR_{\text{N}}I_{\text{c}}/\epsilon_{\text{c}}$ at 50 mK is about
2.3, a factor of 5 smaller than expected. The discrepancy may be
due to the theoretical assumption of perfect interfaces, or due to
the prescence of thermally activated phase slipping in 1-D
proximity-coupled normal wire at low temperatures,~\cite{Tinkham1996} which can
lead to premature switching. In addition, since the length of the normal metal arm is
 comparable to the phase coherence length, the suppression of critical current
 could also be attributed to various dephasing effects~\cite{Golubev2007}.

\begin{figure}
\includegraphics[width=3.1in]{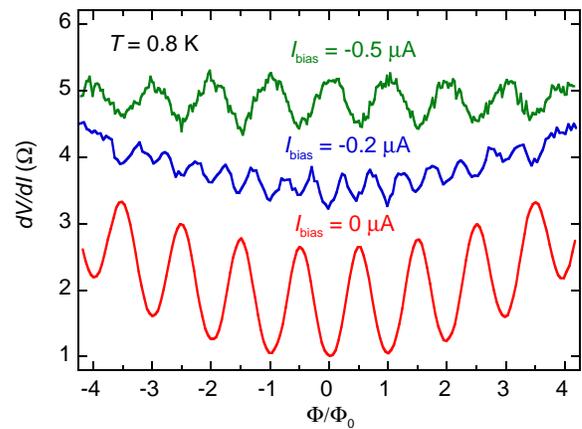}
\caption{(Color online) Magnetoresistance of the top loop in Fig.~1 at 800 mK for
different dc bias currents. The applied flux is measured in units
of superconducting flux quanta, $\Phi_{\text{0}}=h/2e$, corresponding to
a field of 47 G for this device. With increasing bias current, the
oscillation evolves from $h/2e$ periodic oscillations to $h/4e$
periodic oscillations and then to inverted $h/2e$ oscillations.}
\end{figure}

Figure~2 shows the differential resistance as a function of
magnetic flux through the loop at different dc bias currents
($I_{\text{bias}}$). The data is taken at 0.8 K where the normal-metal
loop is in the proximity regime. At zero dc bias, the
magnetoresistance shows large oscillations with a period
corresponding to one superconducting flux quantum $\Phi_{\text{0}}=h/2e$
for the enclosed area of the loop. The amplitude of the
oscillation is about 1.8 $\Omega$, more than $30\%$ of the normal
state resistance. Similar $h/2e$ oscillations have been reported
previously, and were interpreted as the interference of long-range
coherent quasiparticles.~\cite{Petrashov1993,Courtois1996}
However, in this earlier work the proximity effect was thought to be
determined by $T_{\text{Th}}$ and the amplitude of the oscillations by
$T_{\text{Th}}/T$.~\cite{Golubov1997,Courtois1999} Although
 $T_{\text{Th}}$ of our devices is similar to that of these
earlier devices, the amplitude of the observed oscillations is
much larger and the Josephson coupling begins to dominate at a
much higher temperature than $T_{\text{Th}}$.

When a dc bias $I_{\text{bias}}=-0.2$ $\mu$A is added to the ac measurement
current, oscillations in the flux with half the period, i.e., 
$h/4e$, are clearly present (see Fig.2). The $h/4e$ oscillations
here follow a similar quadratic envelope as that of
the $h/2e$ oscillations, and also have their minima at integer
flux quanta, although the amplitude of the oscillations is
smaller. When a higher dc bias $I_{\text{bias}}=-0.5$ $\mu$A is applied,
the $h/2e$ oscillations are recovered with a $\pi$ phase shift
compared to the $h/2e$ oscillations at zero-bias current. The
differential resistance at this bias is close to the normal-state
resistance and the background magnetoresistance is flat.

\begin{figure}
\includegraphics[width=3.1in]{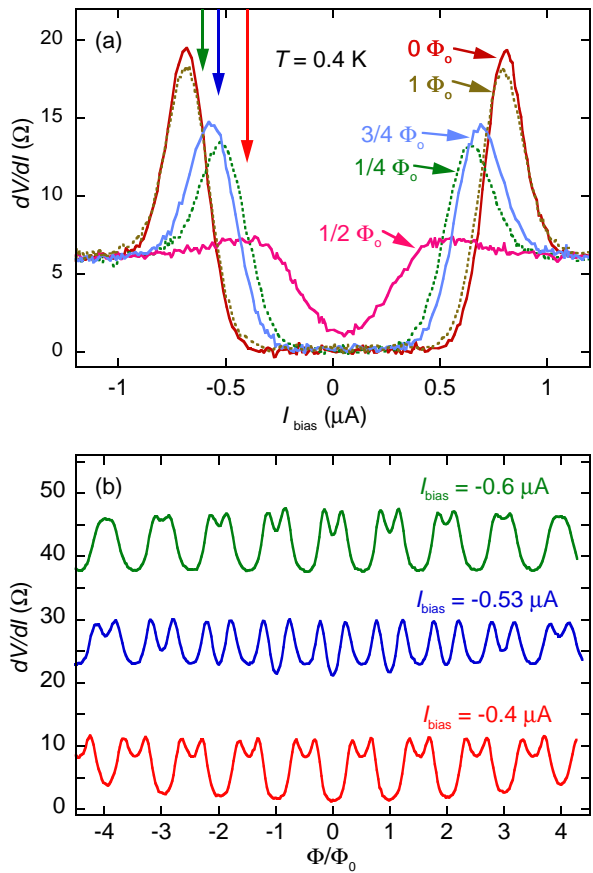}
\caption{(Color online) (a) dc bias current dependence of the differential
resistance at 0.4 K for different applied magnetic fluxes. The
three vertical arrows indicate where two curves cross each other.
(b) From the bottom up, the differential magnetoresistance at
-0.4, -0.53, and -0.6 $\mu$A respectively, showing the evolution to $h/4e$ oscillations. 
The two upper curves are offset by 15 $\Omega$ and 30 $\Omega$ respectively for clarity.}
\end{figure}

To demonstrate how the $h/4e$ oscillations are produced, the
differential resistance as a function of bias current for
different values of magnetic flux through the loop is measured
(Fig.~3(a)).~\cite{note1} At $I_{\text{bias}}=-0.53$ $\mu$A
(marked by the middle vertical arrow) the differential resistance
at zero and one flux quantum is close to that at a half-flux
quantum, and the differential resistance at one-quarter and
three-quarter flux quanta are both higher than that at a
half-flux quantum. Thus the magnetoresistance at
$I_{\text{bias}}=-0.53$ $\mu$A oscillates with half the usual
period, as shown in Fig.~3(b). Note that here the amplitude of
$h/4e$ oscillation can be larger than the normal-state resistance
since the differential resistance is determined by the slope of
the I-V curve. At a slightly lower bias current,
$I_{\text{bias}}=-0.4$ $\mu$A, the $h/4e$ oscillation starts to
appear on top of the $h/2e$ oscillation. At a slightly higher bias
current, $I_{\text{bias}}=-0.6$ $\mu$A, inverted $h/2e$
oscillations replace the $h/2e$ oscillations.

Oscillations with $h/4e$ period have been reported before in
mesoscopic SNS structures~\cite{Petrashov1993} and in
superconducting cylinders.~\cite{Zadorozhny2001a} However, in
these cases no dc bias was intentionally applied, and the
amplitude of the $h/4e$ oscillations was orders of magnitude
smaller. The $h/4e$ oscillations were ascribed to the interplay
between multiple Andreev reflections and inteference in the first
case,~\cite{Petrashov1993} and to an odd number of $\pi$-junctions
in the second case.~\cite{Zadorozhny2001a} The observation of
$h/4e$ oscillations in our own device leads us to another possible
explanation for the oscillations seen previously: as the two arms
of a loop are not perfectly symmetrical, some finite dc voltage
can be induced due to rectification of the ac measurement current
and ac noise~\cite{Dubonos2003,Berger2004,Weiss2000}. This dc
voltage may push the device into a biased regime where $h/4e$
oscillations can be observed as in Fig.~3.

\begin{figure}
\includegraphics[width=3.1in]{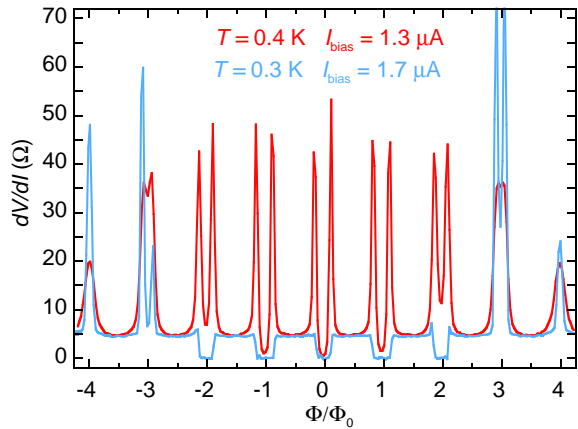}
\caption{(Color online) Differential magnetoresistance at 0.3 K and 0.4 K with dc
bias current set close to the differential resistance peak. The
periodic peaks in the magnetoresistance are mostly absent at 0.3
K, likely due to the peak width at this temperature being smaller
than the measurement current.}
\end{figure}

When the temperature is further lowered, Figure~4 shows that the
differential magnetoresistance peaks become quite sharp, with peak
values more than 10 times the normal state resistance. Although
the $h/2e$ period persists, the shape of the oscillations is no
longer symmetric. For $T=0.3$ K most of the peaks disappear as the
oscillations evolve to become peakless switching from the normal
resistance state to the zero resistance state. The evolution from
sharp peaks to peakless switching was also found for the bias current
dependence of the differential resistance as the
temperature is lowered. In both cases, the disappearance of the
peaks is likely due to the amplitude of the ac measurement current
being larger than the width of the differential resistance peaks
at low temperatures.  For applications this strong dependence of
the differential resistance on the flux may be useful for sensing
small changes in the flux.

In summary, we have measured a dc SNS SQUID in the form of an interrupted
mesoscopic normal-metal loop in contact with two superconducting
electrodes. Unusually large $h/2e$ oscillations, $h/4e$
oscillations, and $h/2e$ oscillation with a $\pi$ phase shift were
found. The $h/4e$ oscillation can be readily explained by
considering the bias current dependence of the differential
resistance.  At lower temperatures these oscillations show high,
flux-dependent differential resistance peaks which evolve to
sharp, peakless switching between the normal and superconducting
states. This rapid change of differential resistance may be useful for
flux sensitive measurements.

This work was funded by
the NSF through grant
DMR-0604601.


\begin{thebibliography}{17}
\bibitem{note1}\bibinfo{note}{Data in this figure were measured after thermal 
cycling to room temperature, which leads to reduced critical currents due to degraded S-N interfaces.}
\end{thebibliography}
\end{document}